\documentclass[11pt]{article}
\usepackage{graphicx} 
\usepackage{amssymb}
\usepackage{amsmath}
\usepackage{geometry}

\usepackage{natbib}
\setcitestyle{authoryear,open={(},close={)}} 

\usepackage{setspace}
\begin{document}

\begin{center}

{\Large{\textbf{A dynamical measure of \\algorithmically infused visibility}}}

\bigskip
{\large{Shaojing Sun$^{1}$, Zhiyuan Liu$^{1}$, David Waxman$^{2}$}}
\bigskip

\noindent\textsuperscript{1}School of Journalism, Fudan University, \\400 Guoding Road, Shanghai 200433, PRC.
\bigskip

\noindent\textsuperscript{2}ISTBI, Fudan University, \\220 Handan Road, Shanghai 200433, PRC.

\end{center}









\begin{center}
\noindent\textbf{Abstract}
\end{center}

This work focuses on the nature of visibility in societies where the behaviours of humans and algorithms 
influence each other  - termed algorithmically infused societies. We propose a quantitative measure
of visibility, with implications and applications to an array of disciplines including communication studies, political science, marketing, technology design, and social media analytics. The measure captures the basic characteristics of the visibility of a given topic, in algorithm/AI-mediated communication/social media settings. Topics, when trending, are ranked against each other, and the proposed measure combines the following two attributes of a topic: (i) the amount of time a topic spends at different ranks, and (ii) the different ranks the topic attains. The proposed measure incorporates a tunable parameter, termed the discrimination level, whose value determines the relative weights of the two attributes that contribute to visibility. Analysis of a large-scale, real-time dataset of trending topics, from one of the largest social media platforms, demonstrates that the proposed measure can explain a large share of the variability of the accumulated  views of a topic.
\newline
\textit{Keywords}:  social media, trending topics, views, AI-mediated communication, measurement, algorithm

\bigskip


\newpage
\begin{center}
    \textbf{A Dynamical Measure of Algorithmically Infused Visibility}
\end{center}
Human beings are entering an ``algorithmically infused society" namely a society 
that is increasingly being shaped by the combined behaviours of humans and algorithms/AI \citep{Wagner}. 
Algorithms are not just employed in the selection and distribution of information, 
but are \textit{dynamic actors} that have the effect of organising and shaping behaviour at the level of the individual,
and also at the level of society as a whole. 
As an illustration of this,
\citet{Guess}
recently demonstrated that the feed-ranking algorithms of Facebook and Instagram had a strong influence 
on the experience of users of these two platforms, and a ``reverse-chronologically-ordered feed" dramatically reduced the time users spent, 
and decreased their engagement with content.  It seems that the complex role that algorithms play, in the dissemination of information, requires a rethinking of communication processes that are often entangled with the 
``properties of algorithms, the relationality of algorithms, and the temporality of the materialization of algorithms"
\citep{Laapotti}.

In various aspects of our lives, algorithms are becoming more pervasive, and often affect what is seen, recognized, and even remembered. The role that algorithms play, in the mediation of information, has profound implications for the dissemination of information, public discourse, individual experiences, and societal dynamics. Phenomena that are of growing concern, such as filter bubbles, echo chambers, clickbait, marginalized visibility, to name a few, are all intimately connected with the action of algorithms. It seems fair to say that algorithms are exhibiting an increasing impact on what, to the general public, is visible (and what is invisible). 

\subsection*{Conceptualizing visibility in digital space}
The visibility (seeing) of others might initially be described as involving  ``those who share the same spatial-temporal locale’’ and is reciprocal in the sense that ``we can see others who are within our field of vision, but they can also see us’’  \citep{Thompson}. 
However, advances in communication technology have greatly reshaped our understanding of visibility, and new forms 
and properties of visibility have arisen. In the digital context, visibility is freed from spatial and temporal constraints of the here and now, and involves scenarios such as "there and then" or "there and now". In other words, in a media-saturated society, visibility has become increasingly mediated by communication. Researchers in this area have proposed 
new concepts such as \textit{mediated visibility} and \textit{socially mediated visibility} \citep{Lane}, 
to describe the enriched and evolving nature of visibility. 
\citet{Treem} suggested
that visibility is a kind of affordance, and from this perspective visibility signifies the possibility 
of certain actions (e.g., the likelihood of a message to be viewed), and the higher the visibility, the higher 
the possibility for actions such as clicking or viewing to occur. 
Following this logic, it stands to reason that the more visible a trending topic is, in digital space (like a ranking list), the more likely individual users are to perceive such affordances, and the more they will tend to actualize such affordances by clicking or reading the topic.  

\citet{Flyverbom} proposed the concept of \textit{visibility management} to explain information 
flow and control in digital space. Although these authors primarily defined the concept from an organizational and managerial perspective, the notion of visibility management relates to algorithm-sustained 
recommendation/ranking of, e.g.,  trending topics. We argue that the phenomenon of the ranking of trending topics, in social media, can also be considered visibility management, i.e., a dynamical social process that includes the interplay of social media organizations, technologies, individuals and governments. As such, the dynamic, relational, and 
ambiguous workings of vision speak to the power of related stakeholders as well the management of information flow. 

The meaning of visibility has become even more complicated as algorithms begin to exercise an increased impact on the phenomenon. For example, 
\citet{Rieder} looked into the operation and consequence of the YouTube search algorithm, from a social-technical perspective. These authors argued that ranking algorithms are a complex process that unfolds over time, involving multiple actors such as advertisers, end users, and politicians. As such, ranking algorithms are, in essence, ‘ranking cultures’ that are rendered as evolving processes that modulate visibility and network/coordinate different actors. Visibility, in digital media landscapes, is intricately linked to the attention and engagement of users. Digital platforms are designed to capture and hold users’ attention, and visibility often hinges on the ability to stand out and resonate with an audience in this highly competitive environment \citep{Goldhaber}.
 However, in an algorithm-mediated communication setting, visibility is much harder to define and quantify. The reason is that the algorithm-sustained digital space is not static, but fluid and continually changing. The complex mutual influences of algorithms and users make the space intrinsically dynamical. For example, the trending-topic list on some social media platforms (e.g., Weibo - one of the largest Chinese social media platforms, which is broadly equivalent to Twitter/X) is an ontogenetic space, since it entails both being and becoming - with the listing and ranking of `hot' topics continually changing. As such, measurement of digital visibility needs to take into account the fluid evolving/dynamical nature of the space.

Past research on visibility in digital space suggests that visibility is (i) relational and (ii) temporal. It is relational because a visible topic or event is closely connected to other content, users, or events, as explicated by the Actor Network Theory of LaTour. When it comes to the trending-topic space (e.g., Weibo, Twitter/X), there always seems to be a tension between the ‘promise of visibility’ and the ‘threat of invisibility’: a new topic may enter the list of highly ranked topics and an existing one on the list may exit (effectively be pushed out). It is temporal because visible content does not remain in the spotlight, and hence remain visible, all of the time. Rather, due to the algorithm’s moderation and a platform’s selection, a topic’s status, in the space of topics, fluctuates over time. In developing a measure of visibility, the relational and temporal aspects should be explicitly factored in. As
\citet{Wagner} argued, measurement of algorithmic infused societies needs to improve the match between theoretical constructs and operational measures.

\subsection*{Measuring visibility in algorithmically-infused societies}

    In recent years, social scientists have devoted a significant effort to studying the linkage between visibility
    and algorithm-sustained media. For example, in this context,
    \citet{Bucher}  
    critically examined newsfeed and the EdgeRank algorithm of Facebook in terms of 
\begin{itemize}
    \item \textit{affinity}: the interaction between the content viewer and the creator;
\item \textit{weight}: an interaction with particular content is assigned a weight by the algorithm, 
depending on how important the interaction is judged to be;
\item \textit{time decay}: the recency of an interaction.
\end{itemize}
Despite mounting research on visibility in digital settings, there is a lack of coherent theorising about visibility. Indeed, to date, there is little research on how to empirically measure visibility. As a result, most studies rely on platform-provided indices (e.g., the number of likes) to assess visibility. Recently, 
\citet{Wagner} explicated the value of developing valid measures to assess algorithmically infused societies. According to these authors, opacity, dynamicism, interconnectedness, and heterogeneity often lead to ready-made indices being inadequate to capture algorithm-sustained communication processes. Wagner et al. contended that researchers should integrate theory-driven and data-driven approaches, in order to develop high-quality and transparent measurements of algorithmically infused societies. To echo their call, we believe the development of a \textit{quantitative measure} of visibility, in an algorithm-mediated communication setting, will be of great value. 

There have been previous approaches in measuring changes in ranking (see the work of \citet{Rieder} and
that of \citet{Webber}).
Such approaches (e.g., a rank-biased distance metric) typically give more weight to changes at the top of the list than changes further down. However, to the best of our knowledge, prior metrics have not been tailored to a digital space, like trending topics on social media. Furthermore, sophisticated mathematical approaches often render those metrics formidable, if not inaccessible,  to social science researchers. It is our goal in this work to present an accessible, readily computable, quantitative measure of visibility in an algorithmic ranking space, which captures changes of rank over time. In other words, visibility in the digital space is characterized by multiple changes over time, and a sound measure, in this case, needs to simultaneously reflect changes of rank. Moreover, following the call of Rieder et al., for a strategy of descriptive assemblage — which is focused on analyzing ranking outcomes rather than seeking causality of algorithms, we create our measure by connecting two key attributes, which are the relative ranking position of a topic, and the amount of time it
spends trending at a given rank. The present study is, to the best of our knowledge, the first attempt to develop a theory-based measure of visibility that can readily employ data in the trending-topic space (e.g., Weibo, Twitter/X). Such an academic endeavor may advance theoretical innovations in conceptualizing visibility in a dynamic, systematic, and complex digital context.

The data used in this work came from the Weibo platform (one of the largest 
Chinese social media platforms). The trending-topic space of Weibo presents a list of popular topics which are updated on a minute-by-minute basis. 
In the Supplemental Material, in Figure S1, we give an annotated screenshot of the trending topic space.
The top 50 topics are numbered and presented in a top-down order. Although the Weibo homepage explains that ranking is based on an algorithmic index, named \textit{hotness}, which factors in a number of aspects that include the recency of a topic, the size of the readership, and the importance of a topic, the precise way
the index is calculated, and how the topics are ordered, remain opaque. Generally, we deem lack of transparency a common issue for such algorithmic spaces, as 
\citet{Rieder} discussed, in terms of a descriptive-assemblage approach, to understand ranking culture. 

For our study, we first needed to acquire data. We achieved this by creating a web crawler that captured the 
list of trending topics, every time point (every minute), on Weibo, the Chinese media platform. We obtained 
5,666,450 data entries, that covered 27,012 different topics, for the period 17\textsuperscript{th} 
December 2022 to 8\textsuperscript{th} March 2023. We note that all data collected was not tied to 
information that allows identification of any individual, and all data was publicly available. 

In the quantitative approach that we present next, when we describe a topic as \textit{trending}, it means that 
at the time in question,  the topic has a ranking in the `top fifty' (i.e., its rank has one of the values $1,2,\ldots,50$).  This reflects the data that is publicly available on the platform (Weibo).

\subsection*{Proposing a quantitative measure of visibility}

We now provide a description of the measure of visibility that we propose in this work.
The fundamental objects we deal with are \textit{rank trajectories} or simply \textit{trajectories} 
of different topics. We write a trajectory of a particular topic as

\bigskip

\[
\begin{tabular}
[c]{|cc|}\hline
$Time$ & \multicolumn{1}{|c|}{$Rank$}\\\hline
\multicolumn{1}{|c|}{$t_{1}$} & $R_{1}$\\
\multicolumn{1}{|c|}{$t_{2}$} & $R_{2}$\\
\multicolumn{1}{|c|}{$t_{3}$} & $R_{3}$\\
\multicolumn{1}{|c|}{$\vdots$} & $\vdots$\\
\multicolumn{1}{|c|}{$t_{f}$} & $R_{f}$\\\hline
\end{tabular}
\
\]

\bigskip

\noindent where $t_i$ denotes time (measured in minutes) and $R_i$ is the rank of the topic that is recorded at that time.  
Thus the trajectory of a topic consists of a list of times, and a list of the ranks that the topic attains at these times. 

The time $t_1$ is the first time the topic has a rank in the top $50$ (i.e., it lies in ranks $1,2,\ldots,50$, where $1$ 
corresponds to the highest rank, $2$ corresponds to the next highest rank, ...), and the time $t_f$ is the 
last time the topic has a rank in the top $50$. The above table contains a complete trajectory
of the topic - in the sense that it covers the set of all ranks of a topic, from the first time it enters the top
$50$ (ranks), to the final time it is in the top $50$.

Note that times and the associated ranks are only recorded for ranks in the top $50$. 
Thus in the above table, the times may not be at every minute, and adjacent times in the list may not differ 
by unity. For example, in the above trajectory, we could have the times $t_1=7$, $t_2=8$ and $t_3=10$, 
meaning that the trajectory starts (enters the top $50$) at time seven minutes,
but at the time of nine minutes, 
the rank of the trajectory was outside the top $50$, and was not 
recorded. 

Here, we make some basic assumptions that a measure of visibility 
should have, and then present a quantitative measure of visibility that is compatible with these.

\begin{enumerate}
\item The visibility of a topic should be proportional to the time a topic is
trending, \textit{all other things being equal}. By this we mean given the four trajectories

\[
\begin{tabular}
[c]{cccc}
trajectory $1$ & trajectory $2$ & trajectory $3$ & trajectory $4$\\
&  &  & \\
$
\begin{tabular}
[c]{|cc|}\hline
$Time$ & \multicolumn{1}{|c|}{$Rank$}\\\hline
\multicolumn{1}{|c|}{$1$} & $40$\\
\multicolumn{1}{|c|}{$2$} & $30$\\
\multicolumn{1}{|c|}{$3$} & $50$\\
\multicolumn{1}{|c|}{} & \\
\multicolumn{1}{|c|}{} & \\
\multicolumn{1}{|c|}{} & \\\hline
\end{tabular}
$ & $
\begin{tabular}
[c]{|cc|}\hline
$Time$ & \multicolumn{1}{|c|}{$Rank$}\\\hline
\multicolumn{1}{|c|}{$10$} & $40$\\
\multicolumn{1}{|c|}{$11$} & $40$\\
\multicolumn{1}{|c|}{$12$} & $30$\\
\multicolumn{1}{|c|}{$13$} & $30$\\
\multicolumn{1}{|c|}{$14$} & $50$\\
\multicolumn{1}{|c|}{$15$} & $50$\\\hline
\end{tabular}
$ & $
\begin{tabular}
[c]{|cc|}\hline
$Time$ & \multicolumn{1}{|c|}{$Rank$}\\\hline
\multicolumn{1}{|c|}{$27$} & $40$\\
\multicolumn{1}{|c|}{$28$} & $30$\\
\multicolumn{1}{|c|}{$29$} & $30$\\
\multicolumn{1}{|c|}{$30$} & $50$\\
\multicolumn{1}{|c|}{$31$} & $40$\\
\multicolumn{1}{|c|}{$32$} & $50$\\\hline
\end{tabular}
$ & $
\begin{tabular}
[c]{|cc|}\hline
$Time$ & \multicolumn{1}{|c|}{$Rank$}\\\hline
\multicolumn{1}{|c|}{$27$} & $40$\\
\multicolumn{1}{|c|}{$28$} & $30$\\
\multicolumn{1}{|c|}{$29$} & $30$\\
\multicolumn{1}{|c|}{$30$} & $50$\\
\multicolumn{1}{|c|}{$31$} & $40$\\
\multicolumn{1}{|c|}{$32$} & $20$\\\hline
\end{tabular}
$
\end{tabular}
\]

a reasonable measure of visibility will accord trajectories $2$ and $3$ precisely \textit{twice} 
the visibility of trajectory $1$, because irrespective of the order of appearance of the ranks, 
and the times that the topic is trending, trajectories $2$ and $3$ achieve all of the ranks that trajectory $1$ achieves, but trajectories $2$ and $3$ spend 
\textit{twice} as much time at these ranks as trajectory $1$. 

     \item Times of high rank should contribute more than the corresponding times of low rank.
      For example, if there are two topics, whose trajectories have the same length, and they have ranks which
      agree at all except a single time, then the trajectory 
     with the higher rank at this time will generally have a higher visibility. For the trajectories
     given above, trajectory $4$ will generally have a higher rank than trajectory $3$ because although trajectories $3$ and $4$ have the same length, trajectory $4$ has one time point (at minute $32$) where
     its rank exceeds that of trajectory $3$.    
     
\end{enumerate}
To proceed, we introduce a quantity $D$, that we term the 
\textit{discrimination level}, that intuitively captures the way we can distinguish between
between different ranks, as we shall
shortly explain.
 Then in accordance with the two assumptions given above, we propose that the 
\textit{visibility} of a topic, at a discrimination level of $D$, written $V(D)$, is given by
\begin{equation}
    V(D)=\sum_{i=1}^{f}{\frac{1}{R_i^D}} \label{V def}
\end{equation}
where the sum is over all ranks of a topic where its rank has been recorded (i.e.,
lying in the top $50$ ranks). The discrimination level, $D$, takes values that are non-negative ($D\ge0$). 
We shall present results when $D$ is restricted to a finite range, for example
\begin{equation}
    0 \le D \le 3. \label{D range}
\end{equation}

When the discrimination level has the value zero ($D=0$), we say there is \textit{no discrimination}, 
in the sense 
that the corresponding visibility, $V(0)$, measures the total time the topic is trending (i.e., 
$V(0)$ is simply the \textit{time} the topic spends in ranks $1$ to $50$, irrespective of the value of the rank).
For the trajectories $1$, $2$, $3$ and $4$, given above, the value of $V(0)$ is $3$, $6$, $6$, and $6$,
respectively, because these are simply the number of times where the topic is trending.

When, for example, $D=3$, there is a \textit{very high level of discrimination}, in  the sense that every minute that a 
topic spends at rank $1$ contributes 
\textit{eight} times as much as every minute where the topic is at rank $2$, and \textit{twenty seven} times as much as 
every minute where the topic is at rank $3$. 

Psychology studies have shown that humans shift their attention from one location to another when viewing a complex image, given the limited capacity of the human visual system in simultaneously processing multiple cues \citep{Nothdurft}.
That said, a person is more likely to allocate more attention to a salient target in the scene, and much less attention to nonsalient targets. According to past research on saliency ranking in computer vision, a user's ranking of the salience of objects in a scene does not follow a linearly decreasing process, it is plausible that the top object is accorded a much higher weight than those further down the list \citep{Siris}. In the language
of visibility, given in Eq. (\ref{V def}), this might correspond to a somewhat appreciable value of $D$. 

However, even a discrimination level of $D=1$ is somewhat discerning in the sense that for 
the trajectories $1$, $2$, $3$ and $4$, that are 
given above, the value of $V(1)$ is, to four decimal places, 
$0.0783$, $0.1567$, $0.1567$, and $0.1867$, respectively. Thus, at a discrimination level of $D=1$, 
trajectory $4$ has a visibility that exceeds that of trajectories $2$ and $3$ by more than $19\%$, 
due to the occurrence of a single time where trajectory $4$ has a higher rank.

Discrimination levels that are higher than those we have considered (e.g., $D>3$) are possible, 
and these can expose other aspects of a topic, for example, if we allow $D$ to become arbitrarily large ($D\rightarrow\infty$), then the resulting value of the visibility just counts the amount 
of time the topic spends at rank $1$. 

We note that there is a simple geometric interpretation of the visibility, 
$V(D)$. For a given value of the discrimination level, $D$, 
we can transform a trajectory, with rank and hence `height' $R$, at a given time, to a new trajectory, with 
transformed height $1/R^D$, which represents a measure of the \textit{importance} of the rank attained at the given time, since higher ranks are more important than lower ones. The \textit{area} under 
this transformed `importance' trajectory, for the given discrimination level, is precisely our visibility measure.
Thus our visibility measure is the \textit{accumulated importance of a topic over the time it is trending}. 

In Figure (1) we plot the trajectories of two different topics, and in Figure (2) we plot their
visibilities, as a function of discrimination level, $D$.

The two trajectories in Fig. 1 achieve maximum ranks that differ, and have total trending times which differ. 
It is evident from Fig. 2 that at smaller discrimination levels,  $D$, the longer-lived (longer trending) 
trajectory has higher visibility. However, for sufficiently large $D$ ($D \gtrsim 0.37$), the shorter-lived, 
but higher-rank achieving trajectory, has a visibility that exceeds the longer-lived trajectory. 
This indicates that at an appropriately high discrimination level, time spent at higher ranks more than
counters a longer trending time. According to different criteria of visibility, i.e., for different choices of $D$, 
either of the two topics used for figures 1 and 2 may have the higher level of visibility.

\subsection*{Linking the visibility measure to outcomes}

Algorithm-generated ranking of online content has proven to be linked to attention allocation and various behavioral outcomes. For instance, 
\citet{Pan} found that online search engine ranking of tourist destinations has a significant impact on \textit{clickthrough rates} (CTRs). Specifically, the top ranked results collected high CTRs but the rates decreased precipitously with decreasing rank. In reference to previous literature, we reason that visibility, factoring in both ranking and temporality, should be linked to clicks, views, and other behavioral outcomes, even though those outcomes may not solely depend on the impact of visibility. For instance,
\citet{Lu} recently found that  the Chinese government uses clickbait to compete for visibility of propaganda contents on \textit{WeChat} (a major Chinese social media platform that is broadly equivalent to Facebook). These authors found that placing hyperbolic words and 
exclamation marks in the titles had large and statistically significant effects on the number of views of the corresponding articles.

It is of interest to consider whether our proposed measure of visibility links to, explains or predicts certain outcomes. To this end, consider a particular topic, and a given discrimination level, $D$. The  corresponding visibility, $V(D)$, can be calculated from knowledge of the set of ranks the topic attains over time.   There are various
criteria or indices associated with the topic, of which an important one is the \textit{total number of reads of the topic}. With
\begin{equation}
    N_{reads} = \text{total number of reads}
\end{equation}
we give, in Figure (3), a scatter plot of the `number of reads', $ N_{reads}$, against visibility, $V(D)$, using logarithmically scaled axes. Data on more than of $23,000$ topics was used in the plot.
Also plotted in the figure is the best straight line through the data (red).

As we have already stated, the level of discrimination, $D$, does not have a predetermined value, but may be chosen according 
to the particular application of visibility that is envisaged. We can however, look for the value of $D$ that
results in visibility \textit{best explaining} the platform supplied index - here the total number of reads of a topic. 
To search for this value of $D$, we first chose
a finely spaced set of $D$ values, and for each value of $D$ in the set we fit a straight line through the logarithmically transformed data, as we did in 
Figure (3) - in that case, just for $D=0.80$. We determined the value of the `coefficient of variation', $R^2$, for 
each such fitted line, and hence for each such $D$ value in the set. This yields Figure (4), where it is observed that
$R^2$ achieves a maximum value, written $R_{max}^2$,  at an intermediate value of $D$. 
The maximum is well-defined, but moderate in extent. 
We write the value of $D$ corresponding to $R_{max}^2$
as $D_{max}$. Then for any topic, $\log_{10}(V(D_{max}))$ is the logarithm of 
the visibility that best
explains the variation in the logarithm of the total number of reads of the topic.

Up to now we have treated all topics as having an equivalent status. In point of fact, topics may be grouped into \textit{categories} because the trending-topic space on Weibo features a variety of topics. The Weibo platform provides a classification of the trending topics presented in the ranking list. Specifically, the platform classified all the topics into 26 broad categories including, but not limited to, domestic news, fashion, music, animation, sports, etc. 
It should be noted that the distinction between certain categories may not be clear-cut, especially as the platform has not provided a detailed description of the categorization procedure. Also notably, the distribution of topics across the categories is markedly imbalanced, with some categories (e.g., music) featuring several thousand topics whereas others (e.g., film and TV series) feature only a dozen. To further illustrate the validity of our measure, we have repeated the above analysis by comparing the results across topical categories. 

Our results show that the optimal parameter values vary across these categories/subcultures, and hence indicate 
that ranking results are not solely determined by the algorithm adopted, but rather, the outcome is contingent 
on the interplay between the algorithm and the culture it is embedded within. As Rieder et al argued, the modulation
of visibility on social media is more about ``ranking cultures'' than ``ranking algorithms''   \citep{Rieder}.

The results are summarized in Table 1, with a complete list of all 26 categories, along with some of their properties, given in the Supplemental Material, in Table S1. The results show that for trending topics about sports and domestic news, the quantification of
visibility, by our measure, can explain more than 65\% of the variability of the accumulated viewership. By contrast, for trending topics about social news, visibility can explain approximately 15\% of the variability of the total readership. Due to the complex nature of the algorithmic space, we cannot jump to the conclusion that there is a cause-and-effect relationship between visibility and accumulated viewership. However, our results do show a relatively strong and statistically significant positive correlation between visibility of a topic and its total viewership. This provides some evidence that visibility is a kind of 
affordance for certain actions, such as clicking and reading in this case.

We draw attention to the fact that different categories of topics have different values of the maximum discrimination
level, $D_{max}$. Some topics, such as those concerned with animation or social news, have values of $D_{max}$, 
that are as high as $1.80$. By contrast humor related topics have a $D_{max}$ of around $0.90$. These findings 
suggest that for topics such as animation, predicting the accumulated viewership is more connected with rank rather 
than the time spent at high ranks. By comparison, for topics such as humor, differences in rank are not as influential
as the time spent in the top $50$.

\subsection*{Discussion}

In this study, we have proposed a quantitative measure to gauge visibility in the context of social media trending topics. The creation of the measure is rooted in theorizing visibility in the digital and especially algorithm-infused context. Our measure aligns with the relational, temporal, and dynamic nature of visibility in the digital context. By taking into consideration the relative position of each topic’s rankings across time, we have proposed a measure with a discrimination parameter whose value is not set, but instead varies by research scenarios and needs. Such a maneuver is premised on the assumption that the visibility of an issue or entity is fluid and mutable in the digital space, because it is constituted by the social, technological, discursive, and material processes \citep{Orlikowski}.

In recent years, researchers have investigated the relationship between visibility and other cognitive/affective/behavioral outcomes from diverse angles. For example, 
\citet{Henderson} departed from Visual Saliency Theory and Cognitive Guidance Theory to investigate how meaning and image saliency influence a person’s attention allocation in viewing real-world scenes. These authors found that despite there being a significant impact of an image’s salience, cognitive relevance plays the dominant functional role in guiding and shaping attention allocation. If we view the trending-topic list as a visual space, it is reasonable to argue that both the visual layout and cognitive relevance of those topics might affect the attention allocation and duration. In this study, we did not consider, in any detail, the semantic meaning of each topic. However, this may be a promising direction of future research.

To better conceptualize visibility, it is necessary to tap into the operational nature and logic of digital settings such as trending topic recommendation and social media discussion. 
\citet{Duguay} demonstrated that the operation of Facebook trending topics is governed by social media logic entailing popularity, programmability, connectivity, and datafication. For instance, this platform utilizes and displays volumes of user data (e.g., number of views) to create the impression that these trending topics are happening in real time. Duguay, furthermore, noted that the four aforementioned elements operate under the influence of an overarching logic (automation), rendering a perception that these trending topics are intervention-free and instantaneous. 

Although our study has adopted a computational approach to craft a measure of visibility, future work should consider using mixed methodology to identify key factors driving or shaping visibility in digital communication. 
\citet{Ellison}, for example, found that on Facebook, user motivations and Facebook feed content are salient predictors of click behavior. The authors distinguished four types of users including indiscriminate clicker (those with high frequency of clicking on social media content but without attentive and extended viewings), engaged clicker (those engaging in clicking and paying close attention to social media content), unengaged lurker (who rarely clicks or views social media content), and engaged lurker (who do not click much but do spend time viewing social media content).

As for the application of the proposed measure, we are agnostic to the values of $D$, but certain values may be adopted if particular applications are in mind (e.g., digital advertising). We can use the value of Dmax to expose differences of topics in different domains. In other words, we can use $D_{max}$ as a partial classification of different categories, which may function as an alternative way to tap into the differences in topics. 

Some weaknesses about the current study should be noted. First, due to the constraints of data collection, we were only able to gather data about the aggregate readings of each topic. If researchers have a chance to collect data about the number of real-time readings across time, that will provide a valuable opportunity to explore the correlations between our measure and user’s actual behavioral outcome over time. Second, the categorization of the topics was derived from the platform-provided meta-data. So, misclassification could be a potential problem. It would be of value to study how measured visibility explains user behaviors across topical domains based on a more rigorous classification scheme. 

Developing quantitative measures of algorithmically-infused societies is a challenging but invaluable task. Future research should continue to explore other concepts related to the digital setting and advance the development of sound measures which can better capture the nature of algorithm-mediated communication.

\newpage

\bibliography{SW}


\newpage

\textbf{Figure 1}
\begin{figure}[!ht]
    \includegraphics[width=\textwidth]{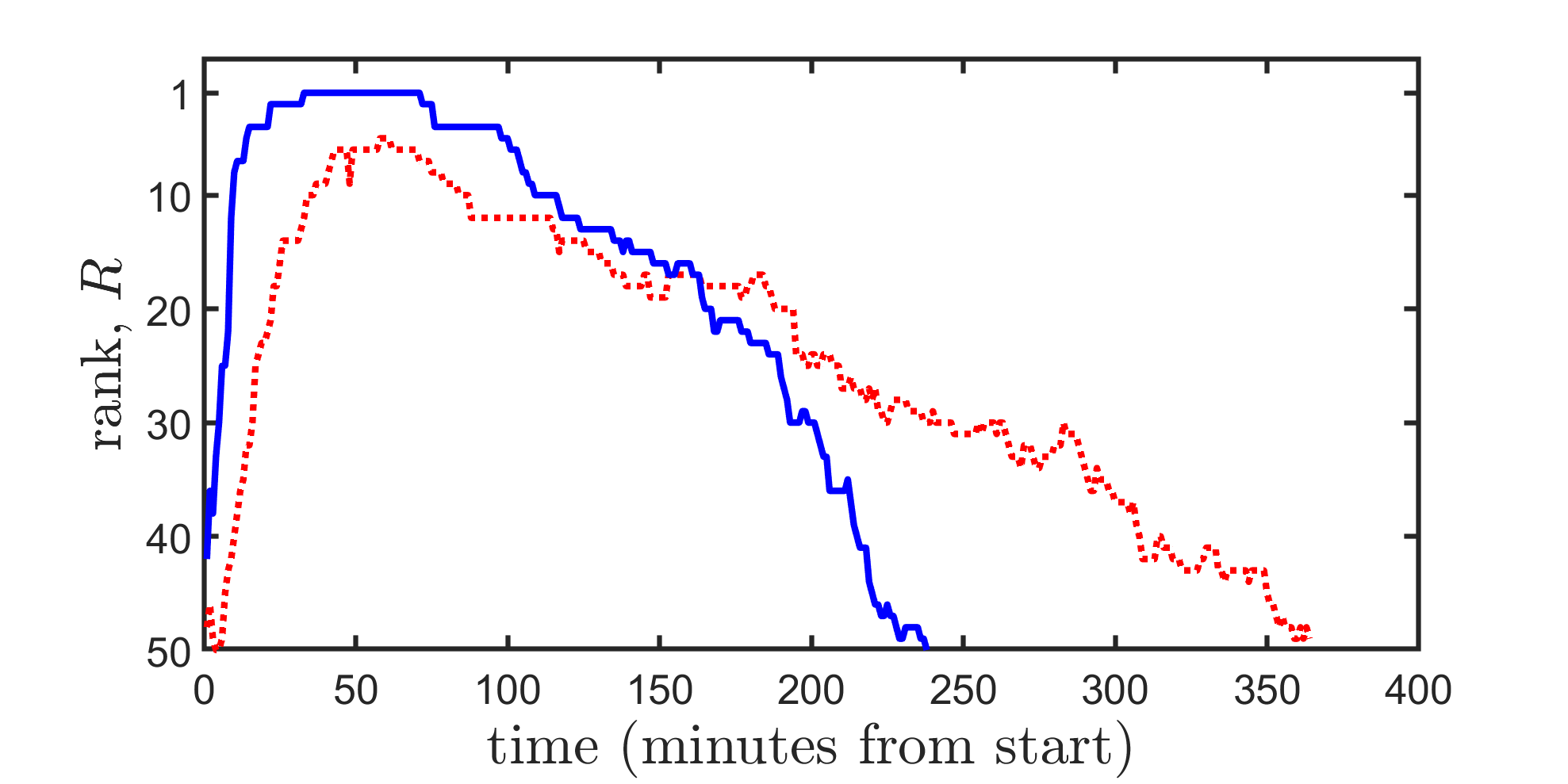}
    \centering
    \caption{Trending trajectories. In the figure, we plot the trajectories (rank, $R$, versus time trending) of two different topics. One topic attains
    a higher maximum rank than the other, while the other topic trends for a longer period of time.}
\end{figure}

\newpage

\textbf{Figure 2}
\begin{figure}[!ht]
    \includegraphics[width=\textwidth]{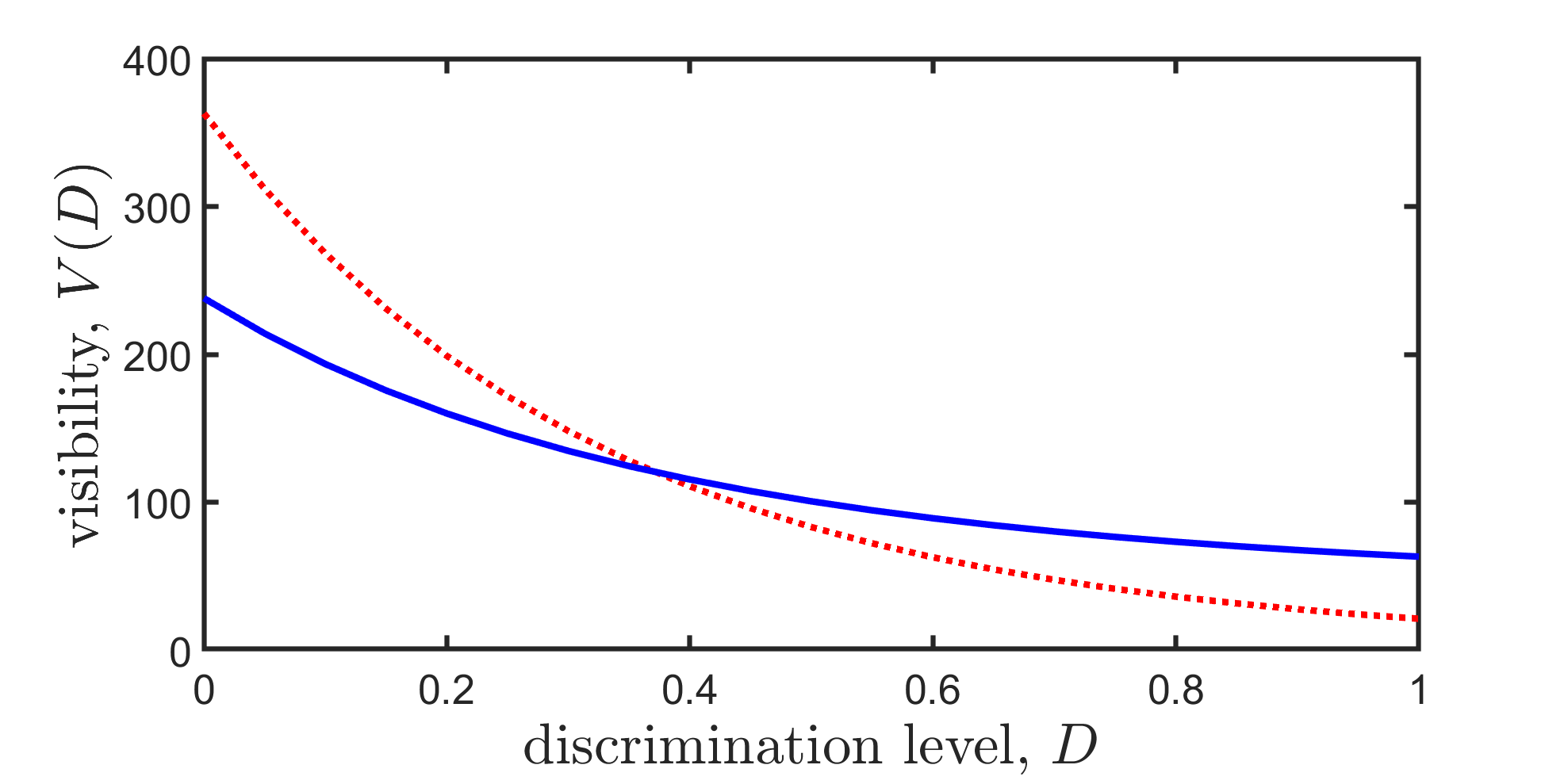}
    \centering
    \caption{Visibility versus discrimination level. In the figure, we plot visibility, $V(D)$, versus discrimination level, $D$, for two topics whose trajectories 
    are given in Figure (1). The two topics spend different amounts of time trending: the values of their 
    visibilities at $D=0$ indicate
    these different values (one topic trends for $363$ minutes, the other for $238$ minutes). 
    For low discrimination levels, the topic with the longer trending time has a higher visibility. 
    However, for sufficiently large $D$ ($D \gtrsim 0.37$), the other topic has a visibility that exceeds the first.
    }
\end{figure}

\newpage

\textbf{Figure 3}
\begin{figure}[!ht]
    \includegraphics[width=\textwidth]{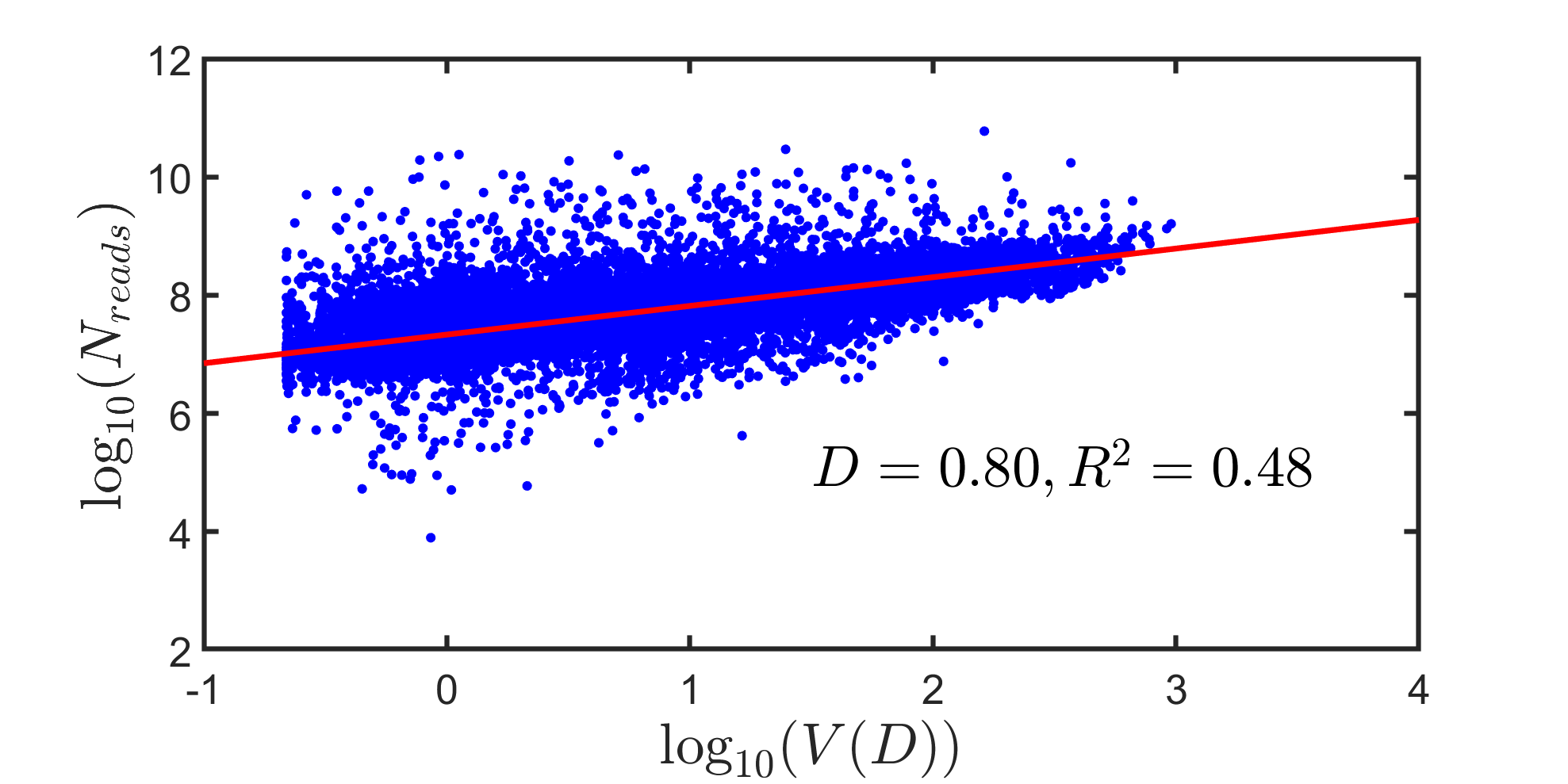}
    \centering
    \caption{Scatter plot showing $N_{reads}$ versus visibility. The figure contains a scatter plot of $\log_{10}( N_{reads} ))$ against $\log_{10}(V(D))$ for points associated 
    with $23,993$ different topics. The figure also contains the best straight line through the data (red).
   A value of the discrimination level of $D=0.8$ was arbitrarily adopted, and the best straight line through the data yielded a value of $R^2$ (coefficient of determination) of $R^2 = 0.48$.}
\end{figure}

\newpage

\textbf{Figure 4}
\begin{figure}[!ht]
    \includegraphics[width=\textwidth]{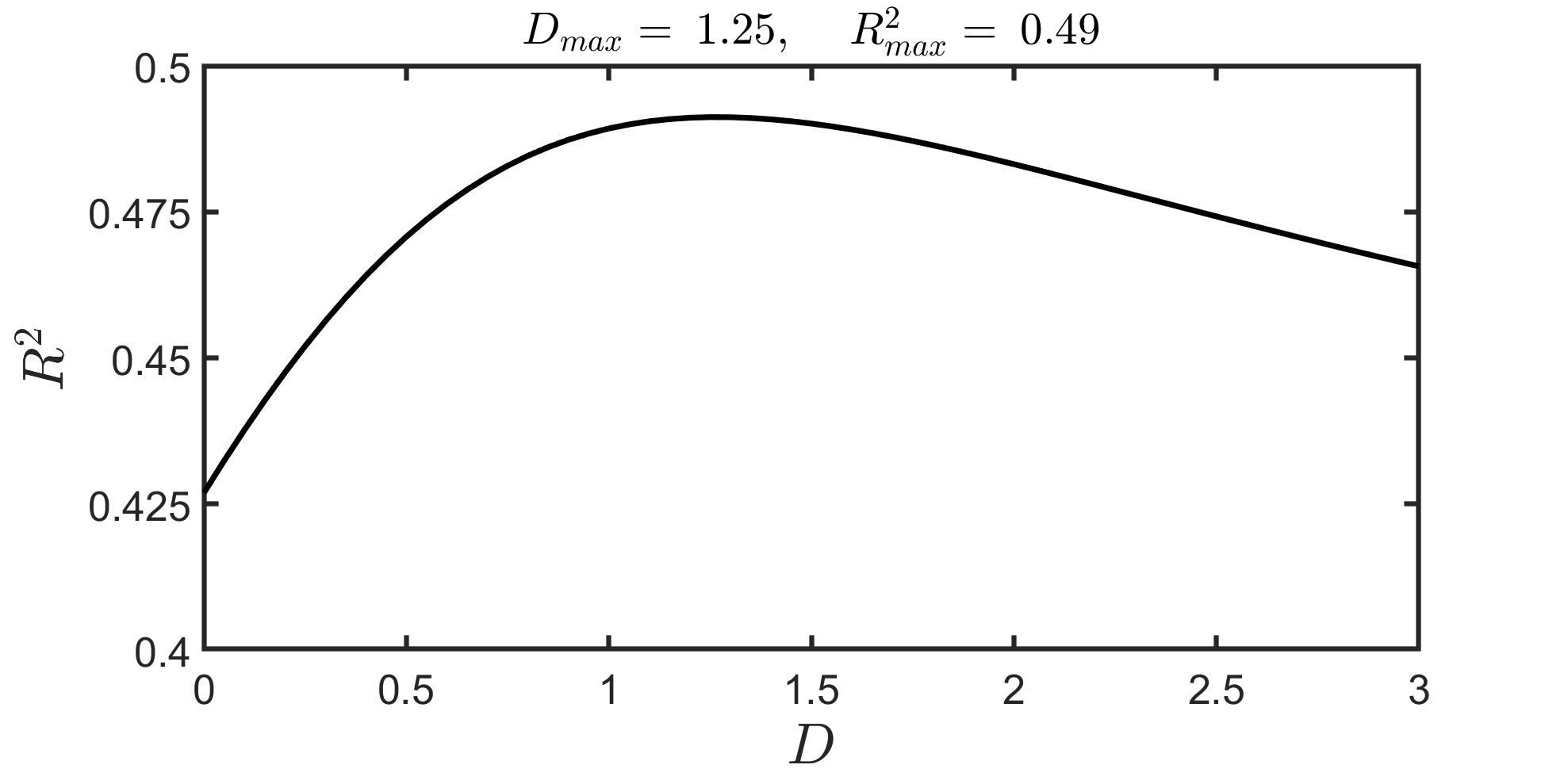}
    \centering
    \caption{Coefficient of determination versus discrimination level. We determined the best straight line of $\log_{10}(N_{reads})$ against $\log_{10}(V(D))$, for $23,993$ different topics, for a set of different values of the discrimination
    level, $D$. The plot contains the $R^2$ values of these lines against  the corresponding values of $D$. We determined the value of $D$ that maximises $R^2$, and  we write these quantities as $D_{max}$ and $R^2_{max}$, respectively.}
\end{figure}

\newpage

\textbf{Table 1}
\[
\begin{tabular}
[c]{|c|c|c|c|}\hline
Category ID & $R_{\max}^{2}$ & $D_{\max}$ & No. of Topics\\\hline
humor & 0.36 & 0.90 & 1549\\\hline
sports & 0.66 & 1.05 & 1209\\\hline
domestic news & 0.68 & 1.05 & 6108\\\hline
music & 0.53 & 1.20 & 4446\\\hline
artists & 0.35 & 1.65 & 2330\\\hline
animation & 0.52 & 1.80 & 1759\\\hline
social news & 0.15 & 1.80 & 1197\\\hline
\end{tabular}
\ \
\]
Table 1 Caption: Properties of major categories. Proportion of variation explained and maximum discrimination level for
\textit{major} topical categories (those containing more than 1000 topics).

\newpage

\setcounter{page}{1}

\begin{center}
    {\Large\textbf{SUPPLEMENTAL MATERIAL}}

    \bigskip
    
    {\Large\textbf{A dynamical measure of \\algorithmically infused visibility}}

    \bigskip

{{Shaojing Sun$^{1}$, Zhiyuan Liu$^{1}$, David Waxman$^{2}$}}

\bigskip

\noindent\textsuperscript{1}School of Journalism, Fudan University, 400 Guoding Road, Shanghai 200433, PRC.

\noindent\textsuperscript{2}ISTBI, Fudan University, 220 Handan Road, Shanghai 200433, PRC.

\end{center}

\bigskip

\bigskip

The Supplemental Material contains two items.

\begin{enumerate}

\item 
Figure S1. An annotated screenshot of the trending topic space, where the top 50 topics are numbered and presented in a top-down order. 

\item

Table S1. This contains a complete list of the 26 broad categories that all topics fall into, including, but not limited to, domestic news, fashion, music, animation, and sports. Alongside the categories in the list are properties of the categories that we have calculated in this work. 

\end{enumerate}

\newpage

\textbf{Figure S1}

\begin{figure}[hbt!]

\label{S1}
    \includegraphics[width=0.75\textwidth]{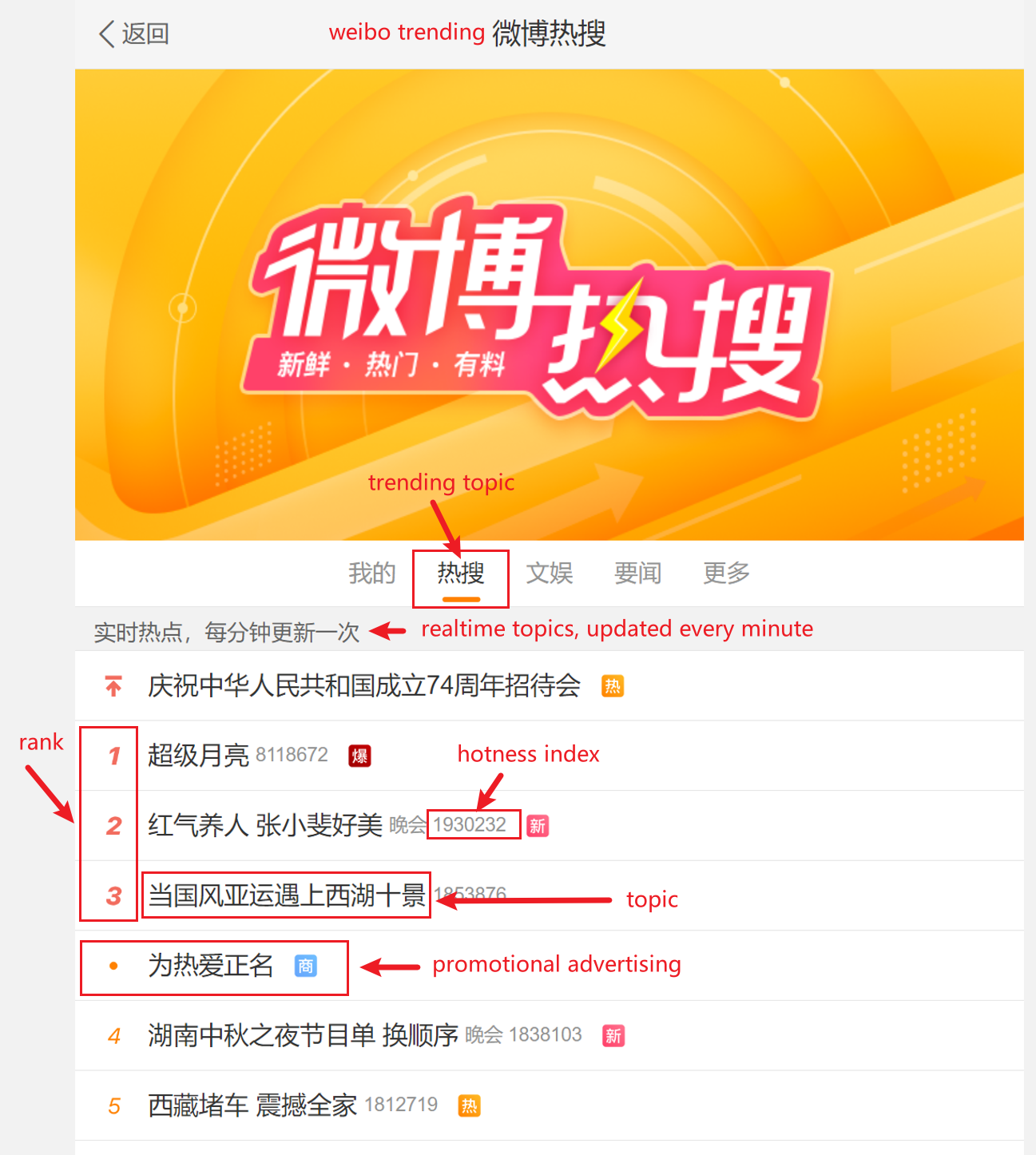}
    \centering
\end{figure}
\noindent{Figure S1: Screenshot of the trending topic space. \\Screenshot from Sina Weibo Mobile, dated 2023/09/29 
Weibo trending url: 
\\https://m.weibo.cn/p/indexcontainerid=106003type\%3D25\%26t\%3D3\%\\26disable\_hot\%3D1\%26filter\_type\%3Drealtimehot}

\newpage
\textbf{Table S1}
\[
\renewcommand\arraystretch{0.56}
\begin{tabular}{|c|l|c|c|c|}
\hline
\multicolumn{2}{|c|}{Category ID} & $R_{\max }^{2}$ & $D_{\max }$ & No. of
Topics \\ \hline
1 & finance & 0.67 & 0.90 & 264 \\ \hline
2 & humor & 0.36 & 0.90 & 1549 \\ \hline
3 & education & 0.41 & 1.20 & 41 \\ \hline
4 & beauty and cosmetics & 0.24 & 3.00 & 54 \\ \hline
5 & sports & 0.66 & 1.05 & 1209 \\ \hline
6 & domestic news & 0.68 & 1.05 & 6108 \\ \hline
7 & social positive energy & 0.28 & 1.50 & 878 \\ \hline
8 & tourism & 0.74 & 1.35 & 108 \\ \hline
9 & electronic games & 0.62 & 1.35 & 136 \\ \hline
10 & film and TV series & 0.29 & 1.35 & 21 \\ \hline
11 & romantic relationships & 0.35 & 1.80 & 905 \\ \hline
12 & animation & 0.52 & 1.80 & 1759 \\ \hline
13 & social news & 0.15 & 1.80 & 1197 \\ \hline
14 & social events & 0.65 & 1.05 & 142 \\ \hline
15 & fashion & 0.68 & 2.40 & 75 \\ \hline
16 & entertainment TV & 0.52 & 1.95 & 945 \\ \hline
17 & online influencer & 0.67 & 1.20 & 256 \\ \hline
18 & unclassified arts & 0.56 & 1.35 & 554 \\ \hline
19 & artists & 0.35 & 1.65 & 2330 \\ \hline
20 & music & 0.53 & 1.20 & 4446 \\ \hline
21 & negative news & 0.42 & 1.05 & 220 \\ \hline
22 & other & 0.33 & 1.50 & 134 \\ \hline
23 & science and tech & 0.23 & 3.00 & 41 \\ \hline
24 & abroad news & 0.40 & 1.05 & 294 \\ \hline
25 & military & 0.73 & 1.20 & 320 \\ \hline
26 & cuisine and food & 0.33 & 1.80 & 222 \\ \hline
\end{tabular}%
\ \ 
\]
\noindent{Table S1: A table of different categories and their properties. \\This table is a list of different categories, along with 
some of their properties. The calculations for $R_{max}^2$ and $D_{max}$ were performed by allowing $D$
to range from $0$ to $3$. The two categories with $D_{max}=3$ in the table (categories $4$ and $23$) 
correspond to no maximum being found in this range. These two categories have a relatively small number of topics.

\end{document}